\documentclass[graybox,natbib,nosecnum]{svmult}
\bibpunct{(}{)}{;}{a}{}{,} 

\usepackage{mathptmx}       
\usepackage{helvet}         
\usepackage{courier}        
\usepackage{type1cm}        

\usepackage{makeidx}         
\usepackage{graphicx, epsfig}        
\usepackage{multicol}        
\usepackage[bottom]{footmisc}
\usepackage[normalem]{ulem}	
\usepackage{hyperref}  

\usepackage{soul}   

\makeindex             

\begin{document}

\title*{The Detectability of Earth's Biosignatures Across Time}
\author{Enric Palle}
\institute{.. \at Instituto de Astrof\'\i sica de Canarias, C. Via Lactea S/N, E-38205 La Laguna, Tenerife, Spain, \email{epalle@iac.es}
}
\maketitle

\abstract{
Over the past two decades, enormous advances in the detection of exoplanets have taken place. Currently, we have discovered hundreds of earth-sized planets, several of them within the habitable zone of their star. In the coming years, the efforts will concentrate in the characterization of these planets and their atmospheres to try to detect the presence of biosignatures. However, even if we discovered a second Earth, it is very unlikely that it would present a stage of evolution similar to the present-day Earth. Our planet has been far from static since its formation about 4.5 Ga ago; on the contrary, during this time, it has undergone multiple changes in it’s atmospheric composition, it’s temperature structure, it’s continental distribution, and even changes in the forms of life that inhabit it. All these changes have affected the global properties of Earth as seen from an astronomical distance. Thus, it is of interest not only to characterize the observables of the Earth as it is today, but also at different epochs. Here we review the detectability of the Earth's globally-averaged properties over time. This includes atmospheric composition and biosignatures, and surface properties that can be interpreted as sings of habitability (bioclues). The resulting picture is that truly unambiguous biosignatures are only detectable for about 1/4 of the Earth's history. The rest of the time we rely on detectable bioclues that can only establish an statistical likelihood for the presence of life on a given planet.
}

\section{Introduction: The Earth in Time}


%
%

In the last few years, we have been able to discover several planets in the super-Earth mass range (e.g \citealt{Udr07b, Cha09, Pep11, Bor12}), some of them lying within, or close to, the habitable zone of their stars (e.g. \citealt{Bor12,Bar13,Ang13}). Even some Earth and Moon-sized planets have been recently announced \citep{Fre12,Mui12,Gil13,Bor13}, and this number is expected to increase in the future. In fact, early statistics have pointed out that around 62\% of the Milky Way's stars might host a super-Earth \citep{Cas12},  while studies from NASA's Kepler mission indicate that about 16.5\% of stars have at least one Earth-size  planet with orbital periods up to 85 days \citep{Fre13}. Particularly interesting are the discoveries of rocky planet around M-type stars \citep{2016Natur.536..437A}, \citep{2017Natur.542..456G}, which due to a better planet/star contrast ratio offer the possibility of exploring their atmospheres in the coming years. Without a doubt, the possibility of finding life will drive the characterization of rocky exoplanets over the coming decades. Still, the search for a truly ''earth-like" planet would imply multiple environmental habitats, the presence of a sizeable atmosphere and complex ecosystems \citep{2016AsBio..16..817S}, and the full enterprise might not be straightforward. 

Because directly imaged extrasolar planets are unlikely to be spatially-resolved, we will have all the planet's information collapsed in a single source of light. Thus, disk-averaged views of Earth are one of  the best way to understand what kind of information one can expect from such type of observations of an Earth analogue. Theoretically, with direct photons from the visible and thermal infrared, and depending on the particular cases, we can characterize a planet in terms of its size, albedo, and, as will be discussed in this paper, its atmospheric gas constituents, total atmospheric column density, clouds, surface properties, land and ocean areas, general habitability, and the possible presence of signs of life. At higher signal-to-noise ratios we will also be able to measure rotation period, weather variability, the presence of land plants, and seasons \citep{2010edp..book.....V}.

However, even if we discovered a second Earth, it is very unlikely that it would present a stage of evolution similar to the present-day Earth.  The Earth has been far from static since its formation about 4.5 Ga ago. On the contrary, during  this time, it has undergone multiple changes in its atmospheric composition,  its temperature structure, its continental distribution, and even changes in the dominant forms of life that  inhabit it. All these changes have affected the global properties of Earth as seen from an astronomical  distance. Thus, it is of interest not only to characterize the observables of the Earth as it is today,  but also at different epochs \citep{Kal07,San12}.

Here we focus on the detectability of its globally-averaged properties over time, without describing the complex underlying geological and biological evolution underneath. For more information on these aspects, the reader is referred to the complementary Olsen et al., chapter in this volume.

%
%
%
%
%
%

%
%

%
%
\section{The Hadean (4.6-4.0 Ga ago) }

The Hadean eon is marked by a post-primary-accretion cataclysmic spike in the number of impacts, commonly referred to as the late heavy bombardment (LHB \citep{2005Natur.435..466G}), and later (about 4.3 Ga ago) the formation of the Earth's oceans \citep{1988JAtS...45.3081A}. At the first stages of planetary formation the earth's surface was a molten layer of rock. 
No crustal rocks are known to have survived from the period 4.55-4.03 Ga \citep{1999CoMP..134....3B}, prior to the Late Heavy Bombardment, although analysis of detrital zircons suggest that continental crust and oceans were present on the early Earth as early as 4.4 Ga ago (\citet{2001Natur.409..178M, 2001Natur.409..175W}).

The Earth's secondary (outgasing) atmosphere was produced during the Hadean eon. The main constituent being Nitrogen, with substantially larger amounts than present day values of  water vapor, $CO_2$, $CH_4$ ad $NH_3$ \citep{1993Sci...259..920K, 2005Sci...308.1014T}, althought the exact amounts and their variability in time remain under debate. Early in the Hadean period $H_2$ and $He$ were also present in the atmosphere, but quickly lost due to atmospheric escape. Nitrogen is still the major component of the present day atmosphere, and it atmospheric abundance is regulated by the life's demands in terms of nutrients \citep{2016AsBio..16..730S}. Although there are some indications that the partial pressure of this gas may have changed over the course of Earth's history, it's overall abundance remains relatively stable \citep{2016AsBio..16..949S}. Figure~\ref{kalten} shows the detectable atmospheric spectral features of Earth along its evolution.

\begin{figure}
\begin{center}
 \includegraphics[width=0.9\textwidth, angle=0]{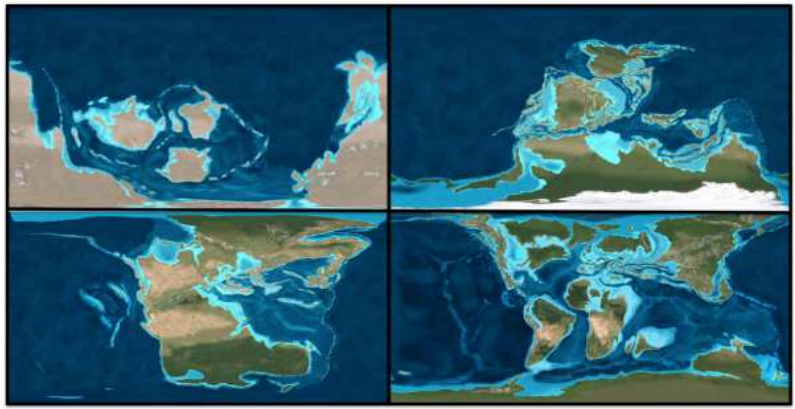}
\caption{Global views of the Earth's continental distribution during the Late Cambrian (500 Ma ago; top left), the Mississippian (340 Ma ago; top right), the Late Triassic (230 Ma ago; bottom left), and the Late Cretaceous (90 Ma ago; bottom right). Courtesy: Ron Blakey, Colorado Plateau Geosystems Inc.}
\label{mapas}
\end{center}
\end{figure}

The LHB is generally regarded as a sterilizing mechanism if life ever arose in the planet during this period, although some authors have argued that there is no plausible situation in which the habitable zone was fully sterilized on Earth, at least since the termination of primary accretion of the planets and the postulated impact origin of the Moon \citep{2009Natur.459..419A}. In fact, organic molecules on the early Earth may have arisen from impact syntheses of bolides colliding in the Hadean oceans \citep{2009NatGe...2...62F}. However, lacking so far scientific evidence of life, the Earth is considered uninhabited during the Hadean, which carries an implicit lack of observable biosignatures.

%
%
\section{The Archean (4.0-2.5 Ga ago)}

It was during the Archean eon that the Earth's magentic field appeared some 3.5-4 billion years ago \citep{2015Sci...349..521T}. It's presence prevented the planet's atmosphere from being stripped away, as the solar wind was 100 times larger than present day values \citep{2010Sci...327.1238T}. It was also during this period that plate building blocks known as cratons, which are essentially giant rock cores, started to come together and rise to the surface \citep{2015PreR..258...48K}. There's evidence of two cratons dated back to as much as 3.5 billion years ago, forming the tiny continent of Vaalbara. It is a supercontinent simply because it was all alone on our planet - any explorers visiting Earth would have seen a single brownish dot against all the blue. Another such craton, Ur, formed roughly 3 billion years ago and actually survived intact as part of larger supercontinents until the break-up of Pangaea only 200 million years ago. The exact distribution of landmasses has an impact on climate trough changes in the averaged bond albedo of the planet \citep{2010Natur.464..744R, 2013JGRD..11810414C}. Still, reasonably detailed maps of Earth's continental distribution are only known for the past 650 million years (see Figure~\ref{mapas}) \footnote{REF http://www.scotese.com/}.

While controversial, the first evidence of life is at 3.8 Ga in isotopically light graphite inclusions in apatite from Greenland \citep{Moj96}, and most likely it was non-photosynthetic, although this is still a subject of debate. The earliest photosynthetic life was probably anoxygenic bacteria like purple bacteria \citep{Xio00,Ols06}, utilizing reductants such as H$_{2}$ or H$_{2}$S instead of water. The Archean biosphere has been proposed to be a mix of anoxygenic phototrophs and chemotrophs such as sulfate-reducing bacteria, methanogens, and other anaerobes \citep{Sec10}. The former perform photosynthesis requiring a band gap energy smaller than that needed to split water, such that the photosynthetically active radiation relevant for anoxygenic photosynthetic bacteria can extend into the near-infrared to as long as $\sim$1025 nm \citep{Sch03}. Thus, their color is distinctly different from that of land plants that dominate the Earth today.

The time when microbial mats appeared on the Earth surface is still not clear, but prior to the evolution of algae and land plants on early Earth, photosynthetic microbial mats probably were among the major forms of life on our planet. Microbial mats are found in the fossil record as early as 3.5 billion years ago. Later, when advanced plants and animals evolved, extensive microbial mats became rarer, but they are still presented in our planet in many ecosystems \citep{Sec10}. Even today, they still persist in special environments such as thermal springs, high salinity environments, and sulfur springs.

\begin{figure*}
\begin{center}
\includegraphics[width=0.9\textwidth, angle=0]{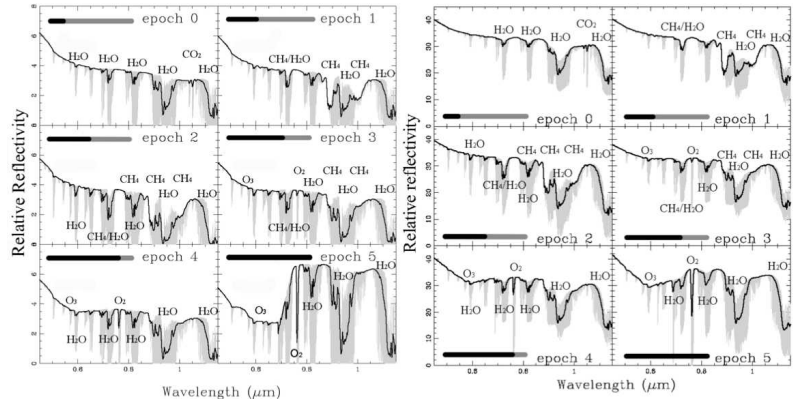} 
\caption{Visible (left) and near-infrared (right) disk-averaged spectra of the Earth for six different geological epochs (3.9, 3.5, 2.4, 2.0, 0.8, and 0.3 Ga ago, from top left to bottom right, respectively).  The models focus on planetary environmental characteristics whose resultant spectral features can be used to imply habitability or the presence of life. These features are generated by $H_2O$, $CO_2$, $CH_4$, $O_2$, $O_3$, $N_2O$, and vegetation-like surface albedos. These epochs exhibit a wide range in abundance for
these molecules, ranging from a $CO_2$-rich early atmosphere, to a $CO_2$/$CH_4$-rich atmosphere around 2 billion years ago,
to a present-day atmosphere. Adapted from \citep{Kal07}
}
\label{kalten}
\end{center}
\end{figure*}

\begin{figure*}
\centering
\includegraphics[width=0.49\textwidth]{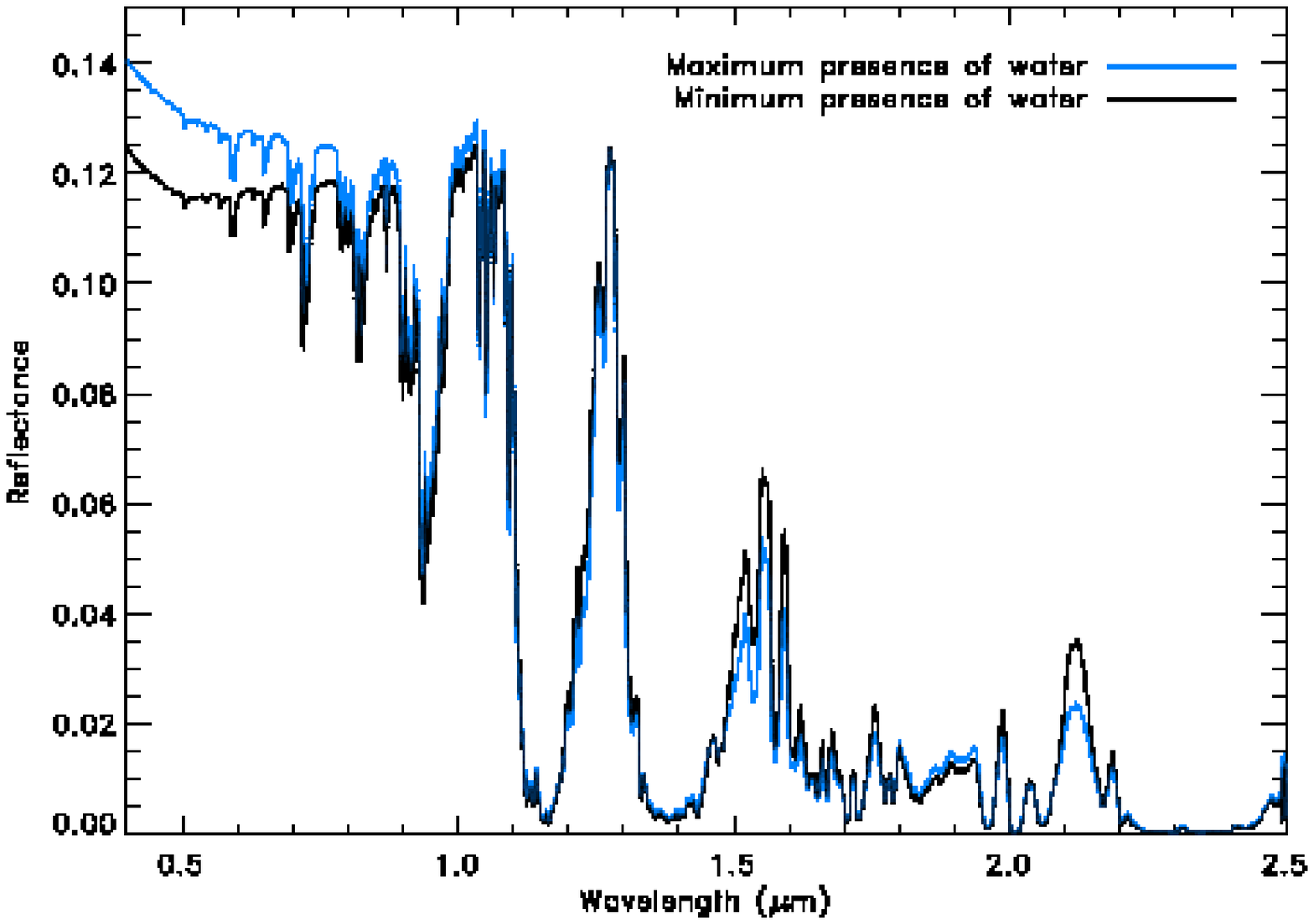}  
\includegraphics[width=0.49\textwidth]{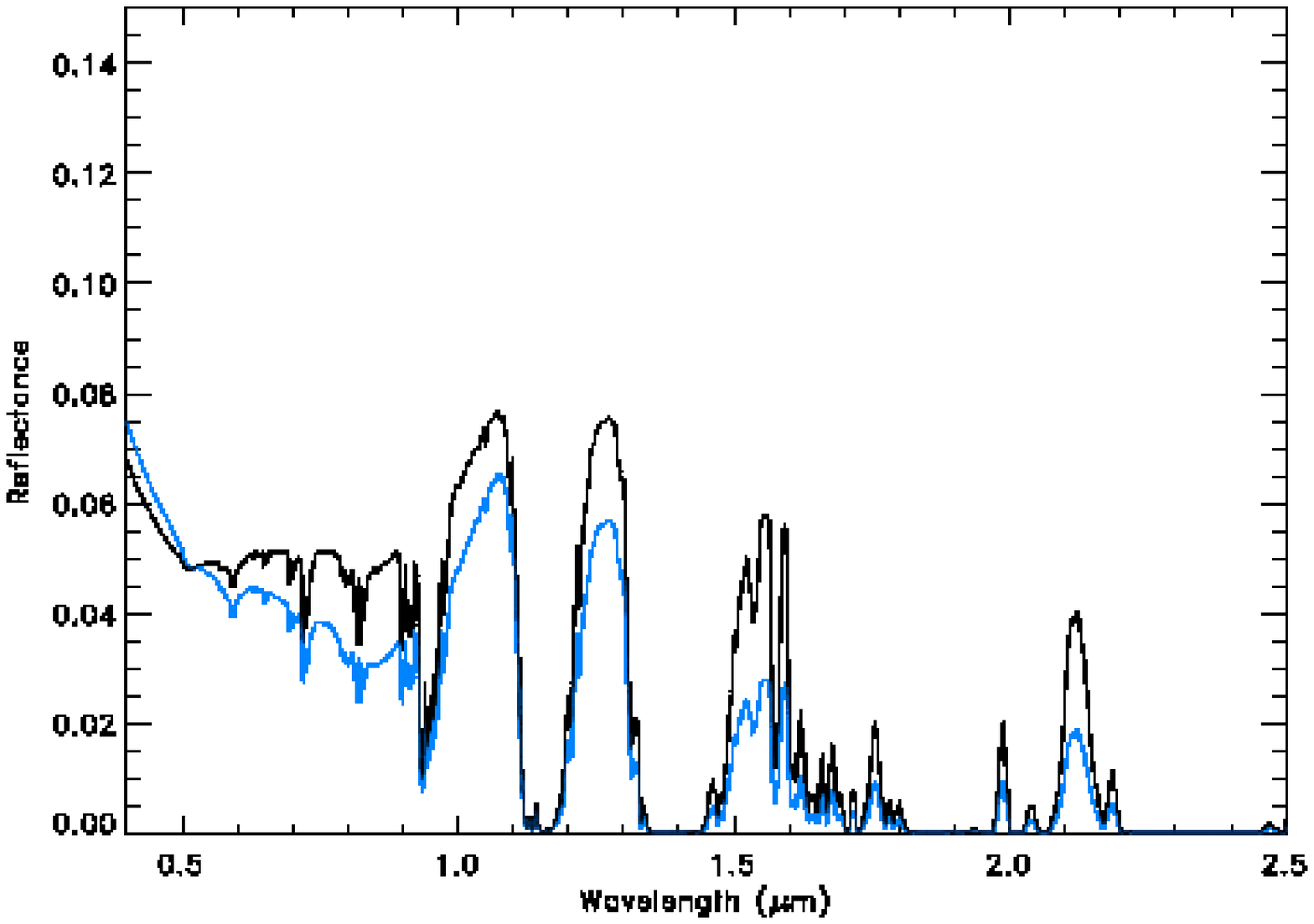} 
\caption{A model of the disk-averaged spectra of the Purple Earth, with the current continental and cloud distribution, but the atmospheric composition corresponding to that of the Archean Earth (3.0 Ga ago). The cloud-free models are shown in the left and the cloudy atmosphere on the right. Continents are assumed to be deserts, but coastal areas are populated with purple bacteria, while oceans are a mixture of water and purple bacteria according to the present-day chlorophyll a distribution. Blue and black models represent whether oceans or continents are predominant from the observer's perspective, respectively. Adapted from \citealt{San13}.}                                                                                              
\label{fig.bac_chl}%
\end{figure*}

During the Archean eon, one of the more widespread life forms on the planet were purple bacteria. These bacteria are photosynthetic microorganisms and can inhabit both aquatic and terrestrial environments, with several species able to live in extreme environments. \citealt{San13} used a radiative transfer model of Earth to simulate the visible and near-IR radiation reflected by our planet, taking into account several scenarios regarding the possible distribution of purple bacteria over continents and oceans. They found that purple bacteria have a reflectance spectrum which has a strong reflectivity increase, similar to the red-edge of leafy plants, although shifted redwards. This feature produces a detectable signal in the disk-averaged spectra of our planet, depending on cloud amount and on purple bacteria concentration/distribution (Figure~\ref{fig.bac_chl}) . They concluded that by using multi-color photometric observations, it is possible to distinguish between an Archean Earth in which purple bacteria inhabit vast extensions of the planet, and a present-day Earth with continents covered by deserts, vegetation or microbial mats. Microbial mats are multilayered sheets of microorganisms generally composed of both Prokaryotes and Eukaryotes, being able to reach a thickness of a few centimeters. \citealt{2015AGUFM.P32B..05P} and \citealt{2013AsBio..13...47H} have also looked at other type of microbial to produce similar edges. This purple Earth scenario would constitute a biosignature similar to the present days' red edge caused by leaf pigmentation.

During the Archean eon, the Sun was about 20\% dimmer than it is today \citep{Gou81,Bah01}, and the atmospheric composition of our planet was completely different to that of present day Earth. At this time, the Earth's atmosphere was likely dominated by $N_{2}$, $CO_{2}$, and water vapor (e.g., \citealt{Wal77,Pin80,Kas93,Kas98}), with little or no free oxygen (Figure~\ref{kalten}). Methane might have also been present as well, helping in the compensation for the reduced solar luminosity (e.g., \citealt{Kie87, Haq08}) and maintaining habitable conditions in the planetary surface. The high concentrations of $CH_4$ at this time would have indicated production by methanogenic bacteria, but $CH_4$ could have been produced abiotically as well \citep{2002AsBio...2..153D}. Thus, observed from an astronomical distance, our planet's atmospheric composition would look like a promising place to search for life, but no conclusive evidence for life could be deduced from the this bulk composition alone  \citealt{2017AsBio..17..287R}. Perhaps with a very in depth characterization of minor species, such as organic sulfur gases abundances \citealt{2011AsBio..11..419D} one could detect the life that produces them, but this might have to wait for very future instrumentation.

The majority of spectral studies that considered Archean Earth and anoxic planets atmospheres have not examined hazes (Meadows 2006; Kaltenegger et al. 2007; Domagal-Goldman et al. 2011) because a haze-rich Archean Earth is expected to be frozen due to the haze’s cooling effects. However, \citealt{2016AsBio..16..873A} used a coupled climate-photochemical-microphysical simulations to show that hazes can cool the planet’s surface by about 20 K, but habitable conditions with liquid surface water could be maintained with a relatively thick haze layer ($\tau$=5 at 200 nm) even with the fainter young sun. They find that optically thicker hazes are self-limiting due to their self-shielding properties, preventing catastrophic cooling of the planet, and could even enhance planetary habitability through UV shielding, reducing surface UV flux by about 97\% compared to a haze-free planet, and potentially allowing survival of land-based organisms 2.6-2.7 billion years ago. The haze in Archean Earth's atmosphere modeled by this researchers was strongly dependent on biologically-produced methane, and they propose that the broad UV absorption signature produced by hydrocarbon haze may be a novel type of spectral biosignature on planets with substantial levels of $CO_2$, although detected alone is not an unambiguous sign of life.

%
%
\section{The Proterozoic  (2.5-0.54 Ga ago) }

The major, defining, event in the Protezozoic period is the transition to an oxygenated atmosphere during the Paleoproterozoic. Though oxygen is believed to have been released by photosynthesis as far back as Archean Eon, it could not build up to any significant degree until mineral sinks of unoxidized sulfur and iron had been filled. But $O_2$ alone is not a biosignature. \citet{2017AsBio..17...27G} pointed out that different "states" of $O_2$ could exist for similar biomass output, and that strong geological activity could lead to false negatives for life, since reducing gases could remove  $O_2$ and mask its biosphere over a wide range of conditions. Still the presence of $O_2$ (or $O_3$ as a proxy) in combination with large amounts of $H_2O$ vapor and  $CH_4$  in the Earth's atmosphere provides the only true realistic biosignature susceptible to be used in the search for life on exoplanets, the so called ``triple fingerprint¨. 

The appearance of the first advanced single-celled, eukaryotes and multi-cellular life, roughly coincides with the start of the accumulation of free oxygen \citet{2010Natur.466..100A}. The blossoming of eukaryotes such as acritarchs did not preclude the expansion of cyanobacteria; in fact, stromatolites reached their greatest abundance and diversity during the Proterozoic, peaking roughly 1200 million years ago \citet{2010Natur.463..885B}. The development of advanced plants is believed to have taken place on Earth during the Late Ordovician, about 450 Ma ago, albeit fungi, algae, and lichens may have greened many land areas before then (Gray et al. 1985). 

During the Proterozoic, our planet surface has also been altered by several snowball events \citep{1998Sci...281.1342H}. Apart from their impact on the albedo and climate, the snowball Earth episodes might be linked to life's evolution. One such snowball Earth episodes occurred just before the Cambrian explosion. Another, much earlier and longer snowball episode, the Huronian glaciation, which occurred 2400 to 2100 Ma, may have been triggered by the first appearance of oxygen in the atmosphere, the "Great Oxygenation Event."

The Proterozoic ends with the Cambrian explosion, which opens the Phanerozoic.

%
%
\section{The  (present Earth)}

The present day Earth atmosphere is dominated by $N_2$ at 78\% and a large amount of $O_2$ at 21\%. Other atmospheric species like $CH_4$ or $CO_2$ are well below their previous historical levels, but they have very strong absorption effects, that on top of making them excellent greenhouse gases, also makes them detectable in the Earth's spectrum (Figure~\ref{kalten}). For the first time in geological history, the globally-integrated spectrum of the planet presents a thermodynamical and chemical disequilibrium, which can be associated to the presence of life \citep{1975RSPSB.189..167L, 2016AsBio..16...39K}.

The start of the Phanerozoic also coincides with the development of land planets, giving rise to another detectable signature in the reflectance spectrum.  Some authors have attempted to detect the vegetation red edge through earthshine measurements (\citealt{Arn02, Woo02, Sea05, Mon06, Ham06}), and also using simulations (\citealt{Tin06a, Tin06b, Mon06}).  The red edge is characterized by strong absorption in the visible part of the spectrum  due to the presence of chlorophyll, which contrasts with a sharp increase in reflectance in the NIR  due to scattering from the refractive index difference between cell walls and the surrounding media.  This particular  signature of vegetation has been proposed as a possible biosignature in Earth-like planets  (e.g., \citealt{Sea05,Mon06,Kia07a}). The possibility of detecting hypothetical alien vegetation  on terrestrial planets has also been studied. \citealt{Tin06c} explored the detectability of  exo-vegetation in a planet orbiting an M star, on which vegetation photosynthetic pigments might  show a shifted red edge signature, and  \citealt{Kia07b} conjectured further about rules for pigments  adaptations to other stellar types.

Photometric color observations of the Earth can also reveal some potentially interesting bioclues. \citet{For01} were the first to point out that the light scattered by a terrestrial planet will vary in intensity and colour as the planet rotates, and the resulting light curve will contain information about the planet’s surface and atmospheric properties. 
However, when clouds are added the reflected light curve is not so straightforward to interpret. The real Earth presents a much more muted light curve due to the smoothing effect of clouds, but the overall albedo is higher \citep{Pal08}.
Clouds are common on the solar system planets, and even on satellites with dense
atmospheres. In fact, clouds are also inferred from observations of free-floating substellar
mass objects \citep{2001ApJ...556..872A}.  But on Earth, clouds are continuously forming and
disappearing, covering an average of about 60\% of the Earth’s surface \citet{Ros96}. This feature is unique in the solar system: only the Earth has large-scale cloud patterns that partially cover the planet and change on timescales
comparable to its rotational period. This is because the temperature and pressure
on the Earth’s surface allow for water to change phase with relative ease from solid
to liquid to gas, providing an excellent bioclue. \citet{Pal08} found that scattered light observations of the Earth can
be used to accurately identify the rotation period of the Earth’s surface, because large-scale
time-averaged cloud patterns are tied to the surface features of Earth, such as continents and ocean currents. 

The identification of the rotation rate of an exoplanet will be important for several reasons: to understand the formation mechanisms and dynamical evolution of extrasolar planetary systems,  to recognize exoplanets that have active weather systems, and even to suggest the presence of a significant magnetic field. Furthermore, if the rotation period of an Earth-like planet can be determined accurately, one can then fold time series of photometric or spectroscopic observations to study regional properties of the planet’s surface and/or atmosphere, improving our sensitivity to detect localized biosignatures. 

Based on the Earth case, in the last few years there have been detailed literature works developing complex deconvolution techniques to identify surface features such as oceans and continents on exoplanets, and studying the possibility of reconstructing longitudinally averaged surface and albedo maps of a planet's surface \citep{Kaw10, Kaw11, Fuj12, Cow11, Rob11, Fuj13}. 

Observations of the Earth's reflected/emitted light at ultraviolet, visible and infrared wavelengths have been scrutinized in in search for all possible biosignatures \citep{Woo02,Qiu03,Pal03, Pal04,Tur06,Pal09, Ham06} and to  determine the possibility of classifying the earth's colors \citep{Cro11}.  Even the use of linear polarization and spectro-polarimetry has been studied as a means to detect clouds and biosignatures \citep{Ste12, 2014A&A...562L...5M}.  A review of the present day Earth seen as an exoplanet can be found at \citet{2010edpr.book.....V}, and a thorough review of Earth's biosignatures can be found in \citealt{2002AsBio...2..153D}.  Still, most of these features remain only a future possibility, way below the technical capabilities of future ground and space instrumentation.

%
%
\section{The Future}

The presence of life on Earth is intrinsically tied with the stability and long-term evolution of the Sun. But inevitably with time the Sun will continue its evolution toward the Red Giant phase, and in this process the Earth will slowly loose it's capability to sustain life and might even disappear engulfed by its parent star.  According to \citealt{2013IJAsB..12...99O}, the Earth's surface will become largely uninhabitable between 1.2 - 1.85 Ga from present, depending on latitude, which is consistent with previous estimates of 1.75 Ga for Earth's habitable lifetime by \citep{2013AsBio..13..833R}. In \citealt{2013IJAsB..12...99O} potential refuge environments in the subsurface and at high altitudes are discussed, which could enable a biosphere to exist for up to 2.8 Ga from the present. Such a biosphere would favour unicellular, anaerobic organisms with a tolerance for one or more extreme conditions. \citealt{2014IJAsB..13..229O} evaluated  the productivity of the biosphere during different stages of biosphere decline between 1 Ga and 2.8 Ga from present, using a simple atmosphere-biosphere interaction model to estimate the atmospheric biosignature gas abundances at each stage and to assess the likelihood of remotely detecting them. Over that period, there is a rapid disappearance of $CO_2$ and $O_2$ (and $CO$ and $O_3$) following the onset of runaway ocean evaporation,  and an associated increase in $H_2$ flux from increased photo-dissociation. A large increase in $CH_4$ is associated with the decay of organic matter for the extreme case of rapid extinction.

\subsection{The role of Intelligence}

The previous section analyzed the scenarios of natural environmental conditions associated only to the physical evolution of the Sun and Earth system. However, this scenario can change substantially if our planet is able to sustain an intelligent species over geological time scales. Several researchers have already indicated that biosignatures of intelligent/technological species, or technosignatures, can be searched for in exoplanet atmospheres or the planet surroundings. These would include: the search for technological albedo edges, such as that of silicon  which might arise from extensive photovoltaic arrays in the planetary surface \citep{2017arXiv170205500L}, the search for artificial structures or estrange transit shapes \citep{2016ApJ...829L...3W}, or the detecion of artifical components in the atmosphere \citep{2010AsBio..10..121S}, among others. 

While dealing with present day climate change, the concept of geo-engineering is already being developped \citep{2000GeoRL..27.2141G}. While this is an option that does not currently count with general support, it will probably become a must for the species to survive in the same planet on Ga time scales.  Thus, major geo-engineering may be seen in inhabited old planets.  For example, a measurement that we might sought include the detection of an unexpectedly large albedo on an Earth-like planet around a red giant, or the detection of artificial components in its atmosphere, such as CFCs in our own atmosphere.

\begin{table}
\caption{Large-scale major detectable biosignatures across time.}
\label{table1}       
%
%
\begin{tabular}{p{2cm}p{2.4cm}p{2cm}p{4.9cm}}
\hline\noalign{\smallskip}
 & Biosignatures &  & Bioclues  \\
\noalign{\smallskip}\svhline\noalign{\smallskip}
Epoch & Atmospheric & Surface &   \\
\noalign{\smallskip}\svhline\noalign{\smallskip}
Hadean          &   --   &   --   &   --  \\
Archean         &   --   &   Bacterial edges   &   $H_2O$, $CH_4$, Clouds, Haze periods  \\
Proterozoic    &   Triple fingerprint   &   Bacterial/Red edges    &  $H_2O$, $CH_4$, Clouds    \\
Phanerozoic   &   Triple fingerprint  &   Red edge &   $H_2O$, $CH_4$, Clouds    \\
Future            &   --   &   --   &   $H_2O$, $CH_4$, Clouds,  \\
Future (+intelligence) &   Triple fingerprint   &   Red edge   &   $H_2O$, $CH_4$, Clouds, Albedo, Techosignatures  \\
\noalign{\smallskip}\hline\noalign{\smallskip}
\end{tabular}
\end{table}

\section{The path to finding life in the galaxy}

One needs to use great caution when trying to predict how to best undertake the search for life in our galaxy, as we are probably going to face unexpected realities. The discovery of exoplanets has revealed a much richer variety of planetary types and planetary system architectures than previously thought. 
It is possible that all life in those planets is very similar to that of Earth, that there is an enormous diversity we are not yet prepared to grasp, or that life is indeed rare, which will make it very difficult to find. Still, we are at the infancy of our search and, at the risk of initially missing parts of the picture, we need to start with some know facts, by taking the Earth as a benchmark for our efforts.

\begin{figure}
\begin{center}
\includegraphics[width=0.95\textwidth, angle=0]{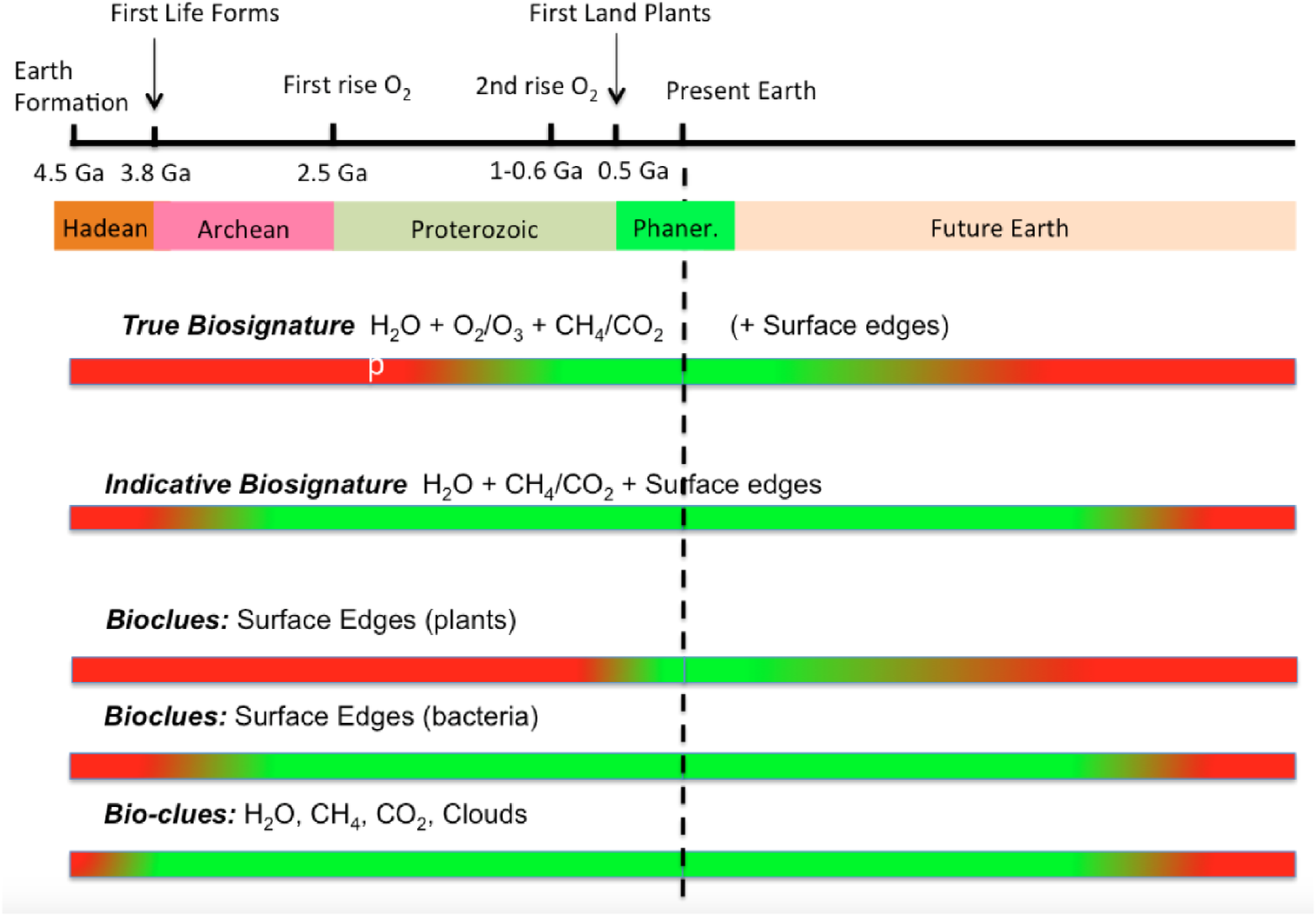} 
\caption{Major detectable atmospheric and surface bioclues and biosignatures along Earth's past, present and future history. The top panel indicates the timeline from Earth's formation into the future, with the geological eons and major events indicated. The several horizontal bars indicate over which period in time a bioclue or a bioignature can be detectable (green) in the globally-averaged spectrum of the Earth or not (red). The vertical dotted line marks the present Earth.  }
\label{megafig}
\end{center}
\end{figure}

It is here useful to make a distinction between what constitutes a `bioclue" and what is a proper biosignature. I use the term bioclue for each piece of information that reveals that the planet might be inhabited, but is not a definitive signs of life. A bioclue for example is the detection of water vapor in the planet's atmosphere, the detection of continental or cloud patters, or even the detection of oxygen, all steps in the right direction, but without offering final probe of life's existence. In Table~\ref{table1} we summarize the major observable bioclues and biosignatures at each geological time on Earth. Many biosignatures have been proposed in the literature, from detection of atmospheric species with only trace amounts to detecting circular polarimetry from biota, but realistically there are very little chances to detect such signatures for many decades to come. So I considered here only those that imply a world-wide signal and have a chance of detection in the next 20-50 years with currently proposed instrumentation and techniques.  
Figure~\ref{megafig} summarizes the detectability of each biosignatures along time.

Table~\ref{table1} and  Figure~\ref{megafig} do not give us a very optimistic framework for the search of life. Over all Earth's history, only during the Phanerozoic and part of the Proterozoic there is a true biosignature detectable in the Earth's spectrum. This is the "triple fingerprint" or the disequilibrium marked by the simultaneous presence of $O_2$, $H_2O$ and $CH_4$ (or $CO_2$). This means that for about 3/4 of the Earth's history, depite being inhabited, life could not have been detected from an astronomical distance. Perhaps with a much advanced technology, in lack of $O_2$, the presence of $H_2O$ and $CH_4$ combined with a statistically significant reflectance edge from surface bacteria could lead to a very high probability scenario for the presence of life, but even this scenario is limited to 2/3 of Earth's history. Several bioclues however, like the presence of $H_2O$, clouds or $CH_4$ are present for most of Earth's history. This means that even if life is a widespread phenomena in our galaxy, detecting it will likely involve a search over numerous targets before a tentative detection can be done.

Thus the future scenario for life detection might consist in accumulating a series of bioclues from a range of astronomical observations and instruments, starting from the planet being in the habitable zone, with which a statistical likelihood of inhabitability of the planet is established. The final detection of the biosignatures will probably need a dedicated mission(s) or instrument(s) for the detailed study of a handful of the most promising targets.  



\begin{acknowledgement}

This work is partly financed by the Spanish Ministry of Economics and Competitiveness through projects ESP2014-57495-C2-1-R and ESP2016-80435-C2-2-R of the Spanish
Secretary of State for R\&D\&i (MINECO).
\end{acknowledgement}

\bibliographystyle{spbasicHBexo}  
\bibliography{biblio2} 

\begin{thebibliography}{96}
\providecommand{\natexlab}[1]{#1}
\providecommand{\url}[1]{{#1}}
\providecommand{\urlprefix}{URL }
\expandafter\ifx\csname urlstyle\endcsname\relax
  \providecommand{\doi}[1]{DOI~\discretionary{}{}{}#1}\else
  \providecommand{\doi}{DOI~\discretionary{}{}{}\begingroup
  \urlstyle{rm}\Url}\fi
\providecommand{\eprint}[2][]{\url{#2}}

\bibitem[{{Abe} and {Matsui}(1988)}]{1988JAtS...45.3081A}
{Abe} Y {Matsui} T (1988) {Evolution of an Impact-Generated H$_{2}$O-CO$_{2}$
  Atmosphere and Formation of a Hot Proto-Ocean on Earth.} Journal of
  Atmospheric Sciences 45:3081--3101

\bibitem[{{Abramov} and {Mojzsis}(2009)}]{2009Natur.459..419A}
{Abramov} O {Mojzsis} SJ (2009) {Microbial habitability of the Hadean Earth
  during the late heavy bombardment}. \nat 459:419--422

\bibitem[{{Ackerman} and {Marley}(2001)}]{2001ApJ...556..872A}
{Ackerman} AS {Marley} MS (2001) {Precipitating Condensation Clouds in
  Substellar Atmospheres}. \apj 556:872--884

\bibitem[{{Albani} et~al.(2010){Albani}, {Bengtson}, {Canfield}, {Bekker},
  {Macchiarelli}, {Mazurier}, {Hammarlund}, {Boulvais}, {Dupuy}, {Fontaine},
  {F{\"u}rsich}, {Gauthier-Lafaye}, {Janvier}, {Javaux}, {Ossa},
  {Pierson-Wickmann}, {Riboulleau}, {Sardini}, {Vachard}, {Whitehouse}, and
  {Meunier}}]{2010Natur.466..100A}
{Albani} AE, {Bengtson} S, {Canfield} DE et~al. (2010) {Large colonial
  organisms with coordinated growth in oxygenated environments 2.1Gyr ago}.
  \nat 466:100--104

\bibitem[{{Anglada-Escud{\'e}} et~al.(2013){Anglada-Escud{\'e}}, {Tuomi},
  {Gerlach}, {Barnes}, {Heller}, {Jenkins}, {Wende}, {Vogt}, {Butler},
  {Reiners}, and {Jones}}]{Ang13}
{Anglada-Escud{\'e}} G, {Tuomi} M, {Gerlach} E et~al. (2013) {A
  dynamically-packed planetary system around GJ 667C with three super-Earths in
  its habitable zone}. \aap 556:A126

\bibitem[{{Anglada-Escud{\'e}} et~al.(2016){Anglada-Escud{\'e}}, {Amado},
  {Barnes}, {Berdi{\~n}as}, {Butler}, {Coleman}, {de La Cueva}, {Dreizler},
  {Endl}, {Giesers}, {Jeffers}, {Jenkins}, {Jones}, {Kiraga}, {K{\"u}rster},
  {L{\'o}pez-Gonz{\'a}lez}, {Marvin}, {Morales}, {Morin}, {Nelson}, {Ortiz},
  {Ofir}, {Paardekooper}, {Reiners}, {Rodr{\'{\i}}guez},
  {Rodr{\'{\i}}guez-L{\'o}pez}, {Sarmiento}, {Strachan}, {Tsapras}, {Tuomi},
  and {Zechmeister}}]{2016Natur.536..437A}
{Anglada-Escud{\'e}} G, {Amado} PJ, {Barnes} J et~al. (2016) {A terrestrial
  planet candidate in a temperate orbit around Proxima Centauri}. \nat
  536:437--440

\bibitem[{{Arney} et~al.(2016){Arney}, {Domagal-Goldman}, {Meadows}, {Wolf},
  {Schwieterman}, {Charnay}, {Claire}, {H{\'e}brard}, and
  {Trainer}}]{2016AsBio..16..873A}
{Arney} G, {Domagal-Goldman} SD, {Meadows} VS et~al. (2016) {The Pale Orange
  Dot: The Spectrum and Habitability of Hazy Archean Earth}. Astrobiology
  16:873--899

\bibitem[{{Arnold} et~al.(2002){Arnold}, {Gillet}, {Lardi{\`e}re}, {Riaud}, and
  {Schneider}}]{Arn02}
{Arnold} L, {Gillet} S, {Lardi{\`e}re} O, {Riaud} P {Schneider} J (2002) {A
  test for the search for life on extrasolar planets. Looking for the
  terrestrial vegetation signature in the Earthshine spectrum}. \aap
  392:231--237

\bibitem[{{Bahcall} et~al.(2001){Bahcall}, {Pinsonneault}, and {Basu}}]{Bah01}
{Bahcall} JN, {Pinsonneault} MH {Basu} S (2001) {Solar Models: Current Epoch
  and Time Dependences, Neutrinos, and Helioseismological Properties}. \apj
  555:990--1012

\bibitem[{{Barclay} et~al.(2013){Barclay}, {Burke}, {Howell}, {Rowe}, {Huber},
  {Isaacson}, {Jenkins}, {Kolbl}, {Marcy}, {Quintana}, {Still}, {Twicken},
  {Bryson}, {Borucki}, {Caldwell}, {Ciardi}, {Clarke}, {Christiansen},
  {Coughlin}, {Fischer}, {Li}, {Haas}, {Hunter}, {Lissauer}, {Mullally},
  {Sabale}, {Seader}, {Smith}, {Tenenbaum}, {Kamal Uddin}, and
  {Thompson}}]{Bar13}
{Barclay} T, {Burke} CJ, {Howell} SB et~al. (2013) {A Super-Earth-sized Planet
  Orbiting in or Near the Habitable Zone around a Sun-like Star}. \apj 768:101

\bibitem[{{Borucki} et~al.(2012){Borucki}, {Koch}, {Batalha}, {Bryson}, {Rowe},
  {Fressin}, {Torres}, {Caldwell}, {Christensen-Dalsgaard}, {Cochran},
  {DeVore}, {Gautier}, {Geary}, {Gilliland}, {Gould}, {Howell}, {Jenkins},
  {Latham}, {Lissauer}, {Marcy}, {Sasselov}, {Boss}, {Charbonneau}, {Ciardi},
  {Kaltenegger}, {Doyle}, {Dupree}, {Ford}, {Fortney}, {Holman}, {Steffen},
  {Mullally}, {Still}, {Tarter}, {Ballard}, {Buchhave}, {Carter},
  {Christiansen}, {Demory}, {D{\'e}sert}, {Dressing}, {Endl}, {Fabrycky},
  {Fischer}, {Haas}, {Henze}, {Horch}, {Howard}, {Isaacson}, {Kjeldsen},
  {Johnson}, {Klaus}, {Kolodziejczak}, {Barclay}, {Li}, {Meibom}, {Prsa},
  {Quinn}, {Quintana}, {Robertson}, {Sherry}, {Shporer}, {Tenenbaum},
  {Thompson}, {Twicken}, {Van Cleve}, {Welsh}, {Basu}, {Chaplin}, {Miglio},
  {Kawaler}, {Arentoft}, {Stello}, {Metcalfe}, {Verner}, {Karoff}, {Lundkvist},
  {Lund}, {Handberg}, {Elsworth}, {Hekker}, {Huber}, {Bedding}, and
  {Rapin}}]{Bor12}
{Borucki} WJ, {Koch} DG, {Batalha} N et~al. (2012) {Kepler-22b: A 2.4
  Earth-radius Planet in the Habitable Zone of a Sun-like Star}. \apj 745:120

\bibitem[{{Borucki} et~al.(2013){Borucki}, {Agol}, {Fressin}, {Kaltenegger},
  {Rowe}, {Isaacson}, {Fischer}, {Batalha}, {Lissauer}, {Marcy}, {Fabrycky},
  {D{\'e}sert}, {Bryson}, {Barclay}, {Bastien}, {Boss}, {Brugamyer},
  {Buchhave}, {Burke}, {Caldwell}, {Carter}, {Charbonneau}, {Crepp},
  {Christensen-Dalsgaard}, {Christiansen}, {Ciardi}, {Cochran}, {DeVore},
  {Doyle}, {Dupree}, {Endl}, {Everett}, {Ford}, {Fortney}, {Gautier}, {Geary},
  {Gould}, {Haas}, {Henze}, {Howard}, {Howell}, {Huber}, {Jenkins}, {Kjeldsen},
  {Kolbl}, {Kolodziejczak}, {Latham}, {Lee}, {Lopez}, {Mullally}, {Orosz},
  {Prsa}, {Quintana}, {Sanchis-Ojeda}, {Sasselov}, {Seader}, {Shporer},
  {Steffen}, {Still}, {Tenenbaum}, {Thompson}, {Torres}, {Twicken}, {Welsh},
  and {Winn}}]{Bor13}
{Borucki} WJ, {Agol} E, {Fressin} F et~al. (2013) {Kepler-62: A Five-Planet
  System with Planets of 1.4 and 1.6 Earth Radii in the Habitable Zone}.
  Science 340:587--590

\bibitem[{{Bowring} and {Williams}(1999)}]{1999CoMP..134....3B}
{Bowring} SA {Williams} IS (1999) {Priscoan (4.00-4.03Ga) orthogneisses from
  northwestern Canada}. Contributions to Mineralogy and Petrology 134:3--16

\bibitem[{{Buick}(2010)}]{2010Natur.463..885B}
{Buick} R (2010) {Early life: Ancient acritarchs}. \nat 463:885--886

\bibitem[{{Cassan} et~al.(2012){Cassan}, {Kubas}, {Beaulieu}, {Dominik},
  {Horne}, {Greenhill}, {Wambsganss}, {Menzies}, {Williams}, {J{\o}rgensen},
  {Udalski}, {Bennett}, {Albrow}, {Batista}, {Brillant}, {Caldwell}, {Cole},
  {Coutures}, {Cook}, {Dieters}, {Prester}, {Donatowicz}, {Fouqu{\'e}}, {Hill},
  {Kains}, {Kane}, {Marquette}, {Martin}, {Pollard}, {Sahu}, {Vinter},
  {Warren}, {Watson}, {Zub}, {Sumi}, {Szyma{\'n}ski}, {Kubiak}, {Poleski},
  {Soszynski}, {Ulaczyk}, {Pietrzy{\'n}ski}, and {Wyrzykowski}}]{Cas12}
{Cassan} A, {Kubas} D, {Beaulieu} JP et~al. (2012) {One or more bound planets
  per Milky Way star from microlensing observations}. \nat 481:167--169

\bibitem[{{Charbonneau} et~al.(2009){Charbonneau}, {Berta}, {Irwin}, {Burke},
  {Nutzman}, {Buchhave}, {Lovis}, {Bonfils}, {Latham}, {Udry}, {Murray-Clay},
  {Holman}, {Falco}, {Winn}, {Queloz}, {Pepe}, {Mayor}, {Delfosse}, and
  {Forveille}}]{Cha09}
{Charbonneau} D, {Berta} ZK, {Irwin} J et~al. (2009) {A super-Earth transiting
  a nearby low-mass star}. \nat 462:891--894

\bibitem[{{Charnay} et~al.(2013){Charnay}, {Forget}, {Wordsworth}, {Leconte},
  {Millour}, {Codron}, and {Spiga}}]{2013JGRD..11810414C}
{Charnay} B, {Forget} F, {Wordsworth} R et~al. (2013) {Exploring the faint
  young Sun problem and the possible climates of the Archean Earth with a 3-D
  GCM}. Journal of Geophysical Research (Atmospheres) 118:10

\bibitem[{{Cowan} et~al.(2011){Cowan}, {Robinson}, {Livengood}, {Deming},
  {Agol}, {A'Hearn}, {Charbonneau}, {Lisse}, {Meadows}, {Seager}, {Shields},
  and {Wellnitz}}]{Cow11}
{Cowan} NB, {Robinson} T, {Livengood} TA et~al. (2011) {Rotational Variability
  of Earth's Polar Regions: Implications for Detecting Snowball Planets}. \apj
  731:76

\bibitem[{{Crow} et~al.(2011){Crow}, {McFadden}, {Robinson}, {Meadows},
  {Livengood}, {Hewagama}, {Barry}, {Deming}, {Lisse}, and {Wellnitz}}]{Cro11}
{Crow} CA, {McFadden} LA, {Robinson} T et~al. (2011) {Views from EPOXI: Colors
  in Our Solar System as an Analog for Extrasolar Planets}. \apj 729:130

\bibitem[{{Des Marais} et~al.(2002){Des Marais}, {Harwit}, {Jucks}, {Kasting},
  {Lin}, {Lunine}, {Schneider}, {Seager}, {Traub}, and
  {Woolf}}]{2002AsBio...2..153D}
{Des Marais} DJ, {Harwit} MO, {Jucks} KW et~al. (2002) {Remote Sensing of
  Planetary Properties and Biosignatures on Extrasolar Terrestrial Planets}.
  Astrobiology 2:153--181

\bibitem[{{Domagal-Goldman} et~al.(2011){Domagal-Goldman}, {Meadows}, {Claire},
  and {Kasting}}]{2011AsBio..11..419D}
{Domagal-Goldman} SD, {Meadows} VS, {Claire} MW {Kasting} JF (2011) {Using
  Biogenic Sulfur Gases as Remotely Detectable Biosignatures on Anoxic
  Planets}. Astrobiology 11:419--441

\bibitem[{{Ford} et~al.(2001){Ford}, {Seager}, and {Turner}}]{For01}
{Ford} EB, {Seager} S {Turner} EL (2001) {Characterization of extrasolar
  terrestrial planets from diurnal photometric variability}. \nat 412:885--887

\bibitem[{{Fressin} et~al.(2012){Fressin}, {Torres}, {Rowe}, {Charbonneau},
  {Rogers}, {Ballard}, {Batalha}, {Borucki}, {Bryson}, {Buchhave}, {Ciardi},
  {D{\'e}sert}, {Dressing}, {Fabrycky}, {Ford}, {Gautier}, {Henze}, {Holman},
  {Howard}, {Howell}, {Jenkins}, {Koch}, {Latham}, {Lissauer}, {Marcy},
  {Quinn}, {Ragozzine}, {Sasselov}, {Seager}, {Barclay}, {Mullally}, {Seader},
  {Still}, {Twicken}, {Thompson}, and {Uddin}}]{Fre12}
{Fressin} F, {Torres} G, {Rowe} JF et~al. (2012) {Two Earth-sized planets
  orbiting Kepler-20}. \nat 482:195--198

\bibitem[{{Fressin} et~al.(2013){Fressin}, {Torres}, {Charbonneau}, {Bryson},
  {Christiansen}, {Dressing}, {Jenkins}, {Walkowicz}, and {Batalha}}]{Fre13}
{Fressin} F, {Torres} G, {Charbonneau} D et~al. (2013) {The False Positive Rate
  of Kepler and the Occurrence of Planets}. \apj 766:81

\bibitem[{{Fujii} and {Kawahara}(2012)}]{Fuj12}
{Fujii} Y {Kawahara} H (2012) {Mapping Earth Analogs from Photometric
  Variability: Spin-Orbit Tomography for Planets in Inclined Orbits}. \apj
  755:101

\bibitem[{{Fujii} et~al.(2013){Fujii}, {Turner}, and {Suto}}]{Fuj13}
{Fujii} Y, {Turner} EL {Suto} Y (2013) {Variability of Water and Oxygen
  Absorption Bands in the Disk-integrated Spectra of Earth}. \apj 765:76

\bibitem[{{Furukawa} et~al.(2009){Furukawa}, {Sekine}, {Oba}, {Kakegawa}, and
  {Nakazawa}}]{2009NatGe...2...62F}
{Furukawa} Y, {Sekine} T, {Oba} M, {Kakegawa} T {Nakazawa} H (2009)
  {Biomolecule formation by oceanic impacts on early Earth}. Nature Geoscience
  2:62--66

\bibitem[{{Gebauer} et~al.(2017){Gebauer}, {Grenfell}, {Stock}, {Lehmann},
  {Godolt}, {von Paris}, and {Rauer}}]{2017AsBio..17...27G}
{Gebauer} S, {Grenfell} JL, {Stock} JW et~al. (2017) {Evolution of Earth-like
  Extrasolar Planetary Atmospheres: Assessing the Atmospheres and Biospheres of
  Early Earth Analog Planets with a Coupled Atmosphere Biogeochemical Model}.
  Astrobiology 17:27--54

\bibitem[{{Gilliland} et~al.(2013){Gilliland}, {Marcy}, {Rowe}, {Rogers},
  {Torres}, {Fressin}, {Lopez}, {Buchhave}, {Christensen-Dalsgaard}, {Désert},
  {Henze}, {Isaacson}, {Jenkins}, {Lissauer}, {Chaplin}, {Basu}, {Metcalfe},
  {Elsworth}, {Handberg}, {Hekker}, {Huber}, {Karoff}, {Kjeldsen}, {Lund},
  {Lundkvist}, {Miglio}, {Charbonneau}, {Ford}, {Fortney}, {Haas}, {Howard},
  {Howell}, {Ragozzine}, and {Thompson}}]{Gil13}
{Gilliland} RL, {Marcy} GW, {Rowe} JF et~al. (2013) {Kepler-68: Three Planets,
  One with a Density between that of Earth and Ice Giants}. \apj 766:40

\bibitem[{{Gillon} et~al.(2017){Gillon}, {Triaud}, {Demory}, {Jehin}, {Agol},
  {Deck}, {Lederer}, {de Wit}, {Burdanov}, {Ingalls}, {Bolmont}, {Leconte},
  {Raymond}, {Selsis}, {Turbet}, {Barkaoui}, {Burgasser}, {Burleigh}, {Carey},
  {Chaushev}, {Copperwheat}, {Delrez}, {Fernandes}, {Holdsworth}, {Kotze}, {Van
  Grootel}, {Almleaky}, {Benkhaldoun}, {Magain}, and
  {Queloz}}]{2017Natur.542..456G}
{Gillon} M, {Triaud} AHMJ, {Demory} BO et~al. (2017) {Seven temperate
  terrestrial planets around the nearby ultracool dwarf star TRAPPIST-1}. \nat
  542:456--460

\bibitem[{{Gomes} et~al.(2005){Gomes}, {Levison}, {Tsiganis}, and
  {Morbidelli}}]{2005Natur.435..466G}
{Gomes} R, {Levison} HF, {Tsiganis} K {Morbidelli} A (2005) {Origin of the
  cataclysmic Late Heavy Bombardment period of the terrestrial planets}. \nat
  435:466--469

\bibitem[{{Gough}(1981)}]{Gou81}
{Gough} DO (1981) {Solar interior structure and luminosity variations}.
  \solphys 74:21--34

\bibitem[{{Govindasamy} and {Caldeira}(2000)}]{2000GeoRL..27.2141G}
{Govindasamy} B {Caldeira} K (2000) {Geoengineering Earth's radiation balance
  to mitigate CO$_{2}$-induced climate change}. \grl 27:2141--2144

\bibitem[{{Hamdani} et~al.(2006){Hamdani}, {Arnold}, {Foellmi}, {Berthier},
  {Billeres}, {Briot}, {Fran{\c c}ois}, {Riaud}, and {Schneider}}]{Ham06}
{Hamdani} S, {Arnold} L, {Foellmi} C et~al. (2006) {Biomarkers in disk-averaged
  near-UV to near-IR Earth spectra using Earthshine observations}. \aap
  460:617--624

\bibitem[{{Haqq-Misra} et~al.(2008){Haqq-Misra}, {Domagal-Goldman}, {Kasting},
  and {Kasting}}]{Haq08}
{Haqq-Misra} JD, {Domagal-Goldman} SD, {Kasting} PJ {Kasting} JF (2008) {A
  Revised, Hazy Methane Greenhouse for the Archean Earth}. Astrobiology
  8:1127--1137

\bibitem[{{Hegde} and {Kaltenegger}(2013)}]{2013AsBio..13...47H}
{Hegde} S {Kaltenegger} L (2013) {Colors of Extreme Exo-Earth Environments}.
  Astrobiology 13:47--56

\bibitem[{{Hoffman} et~al.(1998){Hoffman}, {Kaufman}, {Halverson}, and
  {Schrag}}]{1998Sci...281.1342H}
{Hoffman} PF, {Kaufman} AJ, {Halverson} GP {Schrag} DP (1998) {A Neoproterozoic
  Snowball Earth}. Science 281:1342

\bibitem[{{Kaltenegger} et~al.(2007){Kaltenegger}, {Traub}, and
  {Jucks}}]{Kal07}
{Kaltenegger} L, {Traub} WA {Jucks} KW (2007) {Spectral Evolution of an
  Earth-like Planet}. \apj 658:598--616

\bibitem[{{Kamber}(2015)}]{2015PreR..258...48K}
{Kamber} BS (2015) {The evolving nature of terrestrial crust from the Hadean,
  through the Archaean, into the Proterozoic}. Precambrian Research 258:48--82

\bibitem[{{Kasting}(1993{\natexlab{a}})}]{1993Sci...259..920K}
{Kasting} JF (1993{\natexlab{a}}) {Earth's early atmosphere}. Science
  259:920--926

\bibitem[{{Kasting}(1993{\natexlab{b}})}]{Kas93}
{Kasting} JF (1993{\natexlab{b}}) {Earth's early atmosphere}. Science
  259:920--926

\bibitem[{{Kasting} and {Brown}(1998)}]{Kas98}
{Kasting} JF {Brown} LL (1998) {The early atmosphere as a source of biogenic
  compounds. In the molecular origin of life}

\bibitem[{{Kawahara} and {Fujii}(2010)}]{Kaw10}
{Kawahara} H {Fujii} Y (2010) {Global Mapping of Earth-like Exoplanets From
  Scattered Light Curves}. \apj 720:1333--1350

\bibitem[{{Kawahara} and {Fujii}(2011)}]{Kaw11}
{Kawahara} H {Fujii} Y (2011) {Mapping Clouds and Terrain of Earth-like Planets
  from Photometric Variability: Demonstration with Planets in Face-on Orbits}.
  \apjl 739:L62

\bibitem[{{Kiang} et~al.(2007{\natexlab{a}}){Kiang}, {Segura}, {Tinetti},
  {Govindjee}, {Blankenship}, {Cohen}, {Siefert}, {Crisp}, and
  {Meadows}}]{Kia07a}
{Kiang} NY, {Segura} A, {Tinetti} G et~al. (2007{\natexlab{a}}) {Spectral
  Signatures of Photosynthesis. II. Coevolution with Other Stars And The
  Atmosphere on Extrasolar Worlds}. Astrobiology 7:252--274

\bibitem[{{Kiang} et~al.(2007{\natexlab{b}}){Kiang}, {Siefert}, {Govindjee},
  and {Blankenship}}]{Kia07b}
{Kiang} NY, {Siefert} J, {Govindjee} {Blankenship} RE (2007{\natexlab{b}})
  {Spectral Signatures of Photosynthesis. I. Review of Earth Organisms}.
  Astrobiology 7:222--251

\bibitem[{{Kiehl} and {Dickinson}(1987)}]{Kie87}
{Kiehl} JT {Dickinson} RE (1987) {A study of the radiative effects of enhanced
  atmospheric CO\_2 and CH\_4 on early Earth surface temperatures}. \jgr
  92:2991--2998

\bibitem[{{Krissansen-Totton} et~al.(2016){Krissansen-Totton}, {Bergsman}, and
  {Catling}}]{2016AsBio..16...39K}
{Krissansen-Totton} J, {Bergsman} DS {Catling} DC (2016) {On Detecting
  Biospheres from Chemical Thermodynamic Disequilibrium in Planetary
  Atmospheres}. Astrobiology 16:39--67

\bibitem[{{Lingam} and {Loeb}(2017)}]{2017arXiv170205500L}
{Lingam} M {Loeb} A (2017) {Natural and Artificial Spectral Edges in
  Exoplanets}. ArXiv e-prints

\bibitem[{{Lovelock}(1975)}]{1975RSPSB.189..167L}
{Lovelock} JE (1975) {Thermodynamics and the Recognition of Alien Biospheres}.
  Proceedings of the Royal Society of London Series B 189:167--180

\bibitem[{{Miles-P{\'a}ez} et~al.(2014){Miles-P{\'a}ez}, {Pall{\'e}}, and
  {Zapatero Osorio}}]{2014A&A...562L...5M}
{Miles-P{\'a}ez} PA, {Pall{\'e}} E {Zapatero Osorio} MR (2014) {Simultaneous
  optical and near-infrared linear spectropolarimetry of the earthshine}. \aap
  562:L5

\bibitem[{{Mojzsis} et~al.(1996){Mojzsis}, {Arrhenius}, {McKeegan}, {Harrison},
  {Nutman}, and {Friend}}]{Moj96}
{Mojzsis} SJ, {Arrhenius} G, {McKeegan} KD et~al. (1996) {Evidence for life on
  Earth before 3,800 million years ago}. \nat 384:55--59

\bibitem[{{Mojzsis} et~al.(2001){Mojzsis}, {Harrison}, and
  {Pidgeon}}]{2001Natur.409..178M}
{Mojzsis} SJ, {Harrison} TM {Pidgeon} RT (2001) {Oxygen-isotope evidence from
  ancient zircons for liquid water at the Earth's surface 4,300Myr ago}. \nat
  409:178--181

\bibitem[{{Monta{\~n}{\'e}s-Rodr{\'{\i}}guez}
  et~al.(2006){Monta{\~n}{\'e}s-Rodr{\'{\i}}guez}, {Pall{\'e}}, {Goode}, and
  {Mart{\'{\i}}n-Torres}}]{Mon06}
{Monta{\~n}{\'e}s-Rodr{\'{\i}}guez} P, {Pall{\'e}} E, {Goode} PR
  {Mart{\'{\i}}n-Torres} FJ (2006) {Vegetation Signature in the Observed
  Globally Integrated Spectrum of Earth Considering Simultaneous Cloud Data:
  Applications for Extrasolar Planets}. \apj 651:544--552

\bibitem[{{Muirhead} et~al.(2012){Muirhead}, {Johnson}, {Apps}, {Carter},
  {Morton}, {Fabrycky}, {Pineda}, {Bottom}, {Rojas-Ayala}, {Schlawin},
  {Hamren}, {Covey}, {Crepp}, {Stassun}, {Pepper}, {Hebb}, {Kirby}, {Howard},
  {Isaacson}, {Marcy}, {Levitan}, {Diaz-Santos}, {Armus}, and {Lloyd}}]{Mui12}
{Muirhead} PS, {Johnson} JA, {Apps} K et~al. (2012) {Characterizing the Cool
  KOIs. III. KOI 961: A Small Star with Large Proper Motion and Three Small
  Planets}. \apj 747:144

\bibitem[{Olson(2006)}]{Ols06}
Olson J (2006) Photosynthesis in the archean era. Photosynth Res 88(2):109--17

\bibitem[{{O'Malley-James} et~al.(2013){O'Malley-James}, {Greaves}, {Raven},
  and {Cockell}}]{2013IJAsB..12...99O}
{O'Malley-James} JT, {Greaves} JS, {Raven} JA {Cockell} CS (2013) {Swansong
  biospheres: refuges for life and novel microbial biospheres on terrestrial
  planets near the end of their habitable lifetimes}. International Journal of
  Astrobiology 12:99--112

\bibitem[{{O'Malley-James} et~al.(2014){O'Malley-James}, {Cockell}, {Greaves},
  and {Raven}}]{2014IJAsB..13..229O}
{O'Malley-James} JT, {Cockell} CS, {Greaves} JS {Raven} JA (2014) {Swansong
  biospheres II: the final signs of life on terrestrial planets near the end of
  their habitable lifetimes}. International Journal of Astrobiology 13:229--243

\bibitem[{{Pall{\'e}} et~al.(2003){Pall{\'e}}, {Goode}, {Yurchyshyn}, {Qiu},
  {Hickey}, {Monta{\~n}{\'e}s Rodriguez}, {Chu}, {Kolbe}, {Brown}, and
  {Koonin}}]{Pal03}
{Pall{\'e}} E, {Goode} PR, {Yurchyshyn} V et~al. (2003) {Earthshine and the
  Earth's albedo: 2. Observations and simulations over 3 years}. J Geophys Res
  (Atmos) 108:4710

\bibitem[{{Palle} et~al.(2004){Palle}, {Goode}, {Montanes-Rodriguez}, and
  {Koonin}}]{Pal04}
{Palle} E, {Goode} PR, {Montanes-Rodriguez} P {Koonin} SE (2004) {Changes in
  Earth's Reflectance over the Past Two Decades}. Science 304:1299--1301

\bibitem[{{Pall{\'e}} et~al.(2008){Pall{\'e}}, {Ford}, {Seager},
  {Monta{\~n}{\'e}s-Rodr{\'{\i}}guez}, and {Vazquez}}]{Pal08}
{Pall{\'e}} E, {Ford} EB, {Seager} S, {Monta{\~n}{\'e}s-Rodr{\'{\i}}guez} P
  {Vazquez} M (2008) {Identifying the Rotation Rate and the Presence of Dynamic
  Weather on Extrasolar Earth-like Planets from Photometric Observations}. \apj
  676:1319--1329

\bibitem[{{Pall{\'e}} et~al.(2009){Pall{\'e}}, {Zapatero Osorio}, {Barrena},
  {Monta{\~n}{\'e}s-Rodr{\'{\i}}guez}, and {Mart{\'{\i}}n}}]{Pal09}
{Pall{\'e}} E, {Zapatero Osorio} MR, {Barrena} R,
  {Monta{\~n}{\'e}s-Rodr{\'{\i}}guez} P {Mart{\'{\i}}n} EL (2009) {Earth's
  transmission spectrum from lunar eclipse observations}. \nat 459:814--816

\bibitem[{{Parenteau} et~al.(2015){Parenteau}, {Kiang}, {Blankenship},
  {Sanrom{\'a}}, {Palle Bago}, {Hoehler}, {Pierson}, and
  {Meadows}}]{2015AGUFM.P32B..05P}
{Parenteau} MN, {Kiang} NY, {Blankenship} RE et~al. (2015) {Global Surface
  Photosynthetic Biosignatures Prior to the Rise of Oxygen}. AGU Fall Meeting
  Abstracts

\bibitem[{{Pepe} et~al.(2011){Pepe}, {Lovis}, {S{\'e}gransan}, {Benz},
  {Bouchy}, {Dumusque}, {Mayor}, {Queloz}, {Santos}, and {Udry}}]{Pep11}
{Pepe} F, {Lovis} C, {S{\'e}gransan} D et~al. (2011) {The HARPS search for
  Earth-like planets in the habitable zone. I. Very low-mass planets around
  <ASTROBJ>HD 20794</ASTROBJ>, <ASTROBJ>HD 85512</ASTROBJ>, and <ASTROBJ>HD
  192310</ASTROBJ>}. \aap 534:A58

\bibitem[{{Pinto} et~al.(1980){Pinto}, {Gladstone}, and {Yung}}]{Pin80}
{Pinto} JP, {Gladstone} GR {Yung} YL (1980) {Photochemical production of
  formaldehyde in earth's primitive atmosphere}. Science 210:183--185

\bibitem[{{Qiu} et~al.(2003){Qiu}, {Goode}, {Pall{\'e}}, {Yurchyshyn},
  {Hickey}, {Monta{\~n}{\'e}s Rodriguez}, {Chu}, {Kolbe}, {Brown}, and
  {Koonin}}]{Qiu03}
{Qiu} J, {Goode} PR, {Pall{\'e}} E et~al. (2003) {Earthshine and the Earth's
  albedo: 1. Earthshine observations and measurements of the lunar phase
  function for accurate measurements of the Earth's Bond albedo}. J Geophys Res
  (Atmos) 108:4709

\bibitem[{{Reinhard} et~al.(2017){Reinhard}, {Olson}, {Schwieterman}, and
  {Lyons}}]{2017AsBio..17..287R}
{Reinhard} CT, {Olson} SL, {Schwieterman} EW {Lyons} TW (2017) {False Negatives
  for Remote Life Detection on Ocean-Bearing Planets: Lessons from the Early
  Earth}. Astrobiology 17:287--297

\bibitem[{{Robinson} et~al.(2011){Robinson}, {Meadows}, {Crisp}, {Deming},
  {A'Hearn}, {Charbonneau}, {Livengood}, {Seager}, {Barry}, {Hearty},
  {Hewagama}, {Lisse}, {McFadden}, and {Wellnitz}}]{Rob11}
{Robinson} TD, {Meadows} VS, {Crisp} D et~al. (2011) {Earth as an Extrasolar
  Planet: Earth Model Validation Using EPOXI Earth Observations}. Astrobiology
  11:393--408

\bibitem[{{Rosing} et~al.(2010){Rosing}, {Bird}, {Sleep}, and
  {Bjerrum}}]{2010Natur.464..744R}
{Rosing} MT, {Bird} DK, {Sleep} NH {Bjerrum} CJ (2010) {No climate paradox
  under the faint early Sun}. \nat 464:744--747

\bibitem[{{Rossow} et~al.(1996){Rossow}, {Walker}, {Beuschel}, and
  {Roiter}}]{Ros96}
{Rossow} WB, {Walker} AW, {Beuschel} DE {Roiter} MD (1996) {International
  Satellite Cloud Climatology Project (ISCCP): documentation of New Cloud
  Datasets}. World Climate Research Programme Report WMO/TD 737 (Geneva,
  Switzerland: World Meteorological Organization

\bibitem[{{Rushby} et~al.(2013){Rushby}, {Claire}, {Osborn}, and
  {Watson}}]{2013AsBio..13..833R}
{Rushby} AJ, {Claire} MW, {Osborn} H {Watson} AJ (2013) {Habitable Zone
  Lifetimes of Exoplanets around Main Sequence Stars}. Astrobiology 13:833--849

\bibitem[{{Sanrom{\'a}} and {Pall{\'e}}(2012)}]{San12}
{Sanrom{\'a}} E {Pall{\'e}} E (2012) {Reconstructing the Photometric Light
  Curves of Earth as a Planet along Its History}. \apj 744:188

\bibitem[{{Sanrom{\'a}} et~al.(2013){Sanrom{\'a}}, {Pall{\'e}}, and
  {Garc{\'{\i}}a Mun{\~o}z}}]{San13}
{Sanrom{\'a}} E, {Pall{\'e}} E {Garc{\'{\i}}a Mun{\~o}z} A (2013) {On the
  Effects of the Evolution of Microbial Mats and Land Plants on the Earth as a
  Planet. Photometric and Spectroscopic Light Curves of Paleo-Earths}. \apj
  766:133

\bibitem[{{Scheer}(2003)}]{Sch03}
{Scheer} H (2003) {The pigments. In Advances in Photosynthesis and Respiration,
  Vol. 13: Light-Harvesting Antennas in Photosynthesis}. B.R. Green and W.W.
  Parson, Dordrecht, The Netherlands, Kluwer Academic Publishers, pp. 29-81

\bibitem[{{Schneider} et~al.(2010){Schneider}, {L{\'e}ger}, {Fridlund},
  {White}, {Eiroa}, {Henning}, {Herbst}, {Lammer}, {Liseau}, {Paresce},
  {Penny}, {Quirrenbach}, {R{\"o}ttgering}, {Selsis}, {Beichman}, {Danchi},
  {Kaltenegger}, {Lunine}, {Stam}, and {Tinetti}}]{2010AsBio..10..121S}
{Schneider} J, {L{\'e}ger} A, {Fridlund} M et~al. (2010) {The Far Future of
  Exoplanet Direct Characterization}. Astrobiology 10:121--126

\bibitem[{{Schulze-Makuch} and {Guinan}(2016)}]{2016AsBio..16..817S}
{Schulze-Makuch} D {Guinan} E (2016) {Another Earth 2.0? Not So Fast}.
  Astrobiology 16:817--821

\bibitem[{{Seager} et~al.(2005){Seager}, {Turner}, {Schafer}, and
  {Ford}}]{Sea05}
{Seager} S, {Turner} EL, {Schafer} J {Ford} EB (2005) {Vegetation's Red Edge: A
  Possible Spectroscopic Biosignature of Extraterrestrial Plants}. Astrobiology
  5:372--390

\bibitem[{{Seckbach} and {Oren}(2010)}]{Sec10}
{Seckbach} J {Oren} A (2010) {Microbial Mats: Modern and Ancient Microorganisms
  in Stratified Systems}. Cellular Origin, Life in Extreme Habitats and
  Astrobiology, Springer London, Limited

\bibitem[{{Sterzik} et~al.(2012){Sterzik}, {Bagnulo}, and {Palle}}]{Ste12}
{Sterzik} MF, {Bagnulo} S {Palle} E (2012) {Biosignatures as revealed by
  spectropolarimetry of Earthshine}. \nat 483:64--66

\bibitem[{{St{\"u}eken}(2016)}]{2016AsBio..16..730S}
{St{\"u}eken} EE (2016) {Nitrogen in Ancient Mud: A Biosignature?} Astrobiology
  16:730--735

\bibitem[{{St{\"u}eken} et~al.(2016){St{\"u}eken}, {Kipp}, {Koehler},
  {Schwieterman}, {Johnson}, and {Buick}}]{2016AsBio..16..949S}
{St{\"u}eken} EE, {Kipp} MA, {Koehler} MC et~al. (2016) {Modeling pN$_{2}$
  through Geological Time: Implications for Planetary Climates and Atmospheric
  Biosignatures}. Astrobiology 16:949--963

\bibitem[{{Tarduno} et~al.(2010){Tarduno}, {Cottrell}, {Watkeys}, {Hofmann},
  {Doubrovine}, {Mamajek}, {Liu}, {Sibeck}, {Neukirch}, and
  {Usui}}]{2010Sci...327.1238T}
{Tarduno} JA, {Cottrell} RD, {Watkeys} MK et~al. (2010) {Geodynamo, Solar Wind,
  and Magnetopause 3.4 to 3.45 Billion Years Ago}. Science 327:1238

\bibitem[{{Tarduno} et~al.(2015){Tarduno}, {Cottrell}, {Davis}, {Nimmo}, and
  {Bono}}]{2015Sci...349..521T}
{Tarduno} JA, {Cottrell} RD, {Davis} WJ, {Nimmo} F {Bono} RK (2015) {A Hadean
  to Paleoarchean geodynamo recorded by single zircon crystals}. Science
  349:521--524

\bibitem[{{Tian} et~al.(2005){Tian}, {Toon}, {Pavlov}, and {De
  Sterck}}]{2005Sci...308.1014T}
{Tian} F, {Toon} OB, {Pavlov} AA {De Sterck} H (2005) {A Hydrogen-Rich Early
  Earth Atmosphere}. Science 308:1014--1017

\bibitem[{{Tinetti} et~al.(2006{\natexlab{a}}){Tinetti}, {Meadows}, {Crisp},
  {Fong}, {Fishbein}, {Turnbull}, and {Bibring}}]{Tin06a}
{Tinetti} G, {Meadows} VS, {Crisp} D et~al. (2006{\natexlab{a}}) {Detectability
  of Planetary Characteristics in Disk-Averaged Spectra. I: The Earth Model}.
  Astrobiology 6:34--47

\bibitem[{{Tinetti} et~al.(2006{\natexlab{b}}){Tinetti}, {Meadows}, {Crisp},
  {Kiang}, {Kahn}, {Fishbein}, {Velusamy}, and {Turnbull}}]{Tin06b}
{Tinetti} G, {Meadows} VS, {Crisp} D et~al. (2006{\natexlab{b}}) {Detectability
  of Planetary Characteristics in Disk-Averaged Spectra II: Synthetic Spectra
  and Light-Curves of Earth}. Astrobiology 6:881--900

\bibitem[{{Tinetti} et~al.(2006{\natexlab{c}}){Tinetti}, {Rashby}, and
  {Yung}}]{Tin06c}
{Tinetti} G, {Rashby} S {Yung} YL (2006{\natexlab{c}}) {Detectability of
  Red-Edge-shifted Vegetation on Terrestrial Planets Orbiting M Stars}. \apjl
  644:L129--L132

\bibitem[{{Turnbull} et~al.(2006){Turnbull}, {Traub}, {Jucks}, {Woolf},
  {Meyer}, {Gorlova}, {Skrutskie}, and {Wilson}}]{Tur06}
{Turnbull} MC, {Traub} WA, {Jucks} KW et~al. (2006) {Spectrum of a Habitable
  World: Earthshine in the Near-Infrared}. \apj 644:551--559

\bibitem[{{Udry} et~al.(2007){Udry}, {Bonfils}, {Delfosse}, {Forveille},
  {Mayor}, {Perrier}, {Bouchy}, {Lovis}, {Pepe}, {Queloz}, and
  {Bertaux}}]{Udr07b}
{Udry} S, {Bonfils} X, {Delfosse} X et~al. (2007) {The HARPS search for
  southern extra-solar planets. XI. Super-Earths (5 and 8 M$\{${{\o}plus}$\}$)
  in a 3-planet system}. \aap 469:L43--L47

\bibitem[{{V{\'a}zquez} et~al.(2010{\natexlab{a}}){V{\'a}zquez}, {Pall{\'e}},
  and {Monta{\~n}{\'e}s Rodr{\'{\i}}guez}}]{2010edp..book.....V}
{V{\'a}zquez} M, {Pall{\'e}} E {Monta{\~n}{\'e}s Rodr{\'{\i}}guez} P
  (2010{\natexlab{a}}) {The Earth as a Distant Planet}.
  \doi{10.1007/978-1-4419-1684-6}

\bibitem[{{V{\'a}zquez} et~al.(2010{\natexlab{b}}){V{\'a}zquez}, {Pall{\'e}},
  and {Rodr{\'{\i}}guez}}]{2010edpr.book.....V}
{V{\'a}zquez} M, {Pall{\'e}} E {Rodr{\'{\i}}guez} PM (2010{\natexlab{b}}) {The
  Earth as a Distant Planet}. \doi{10.1007/978-1-4419-1684-6}

\bibitem[{{Walker}(1977)}]{Wal77}
{Walker} JCG (1977) {Evolution of the atmosphere}

\bibitem[{{Wilde} et~al.(2001){Wilde}, {Valley}, {Peck}, and
  {Graham}}]{2001Natur.409..175W}
{Wilde} SA, {Valley} JW, {Peck} WH {Graham} CM (2001) {Evidence from detrital
  zircons for the existence of continental crust and oceans on the Earth 4.4Gyr
  ago}. \nat 409:175--178

\bibitem[{{Woolf} et~al.(2002){Woolf}, {Smith}, {Traub}, and {Jucks}}]{Woo02}
{Woolf} NJ, {Smith} PS, {Traub} WA {Jucks} KW (2002) {The Spectrum of
  Earthshine: A Pale Blue Dot Observed from the Ground}. \apj 574:430--433

\bibitem[{{Wright} and {Sigurdsson}(2016)}]{2016ApJ...829L...3W}
{Wright} JT {Sigurdsson} S (2016) {Families of Plausible Solutions to the
  Puzzle of Boyajian Star}. \apjl 829:L3

\bibitem[{Xiong et~al.(2000)Xiong, Fischer, Inoue, Nakahara, and Bauer}]{Xio00}
Xiong J, Fischer W, Inoue K, Nakahara M Bauer C (2000) Molecular evidence for
  the early evolution of photosynthesis. Science 289(5485):1724--30

\end{thebibliography}

\end{document}